\documentclass[conference,a4paper,flushend]{iaria} % (based on IEEEtran.cls)
\usepackage{caption}
\usepackage{tabularx}

\title{Selecting Cybersecurity Requirements: Effects of LLM Use and Professional Software Development Experience}
\author{
  \IEEEauthorblockN{%
    Damjan Fujs \orcidlink{0000-0002-6357-8569}, Damjan Vavpotič \orcidlink{0000-0002-8039-7110}, Tomaž Hovelja \orcidlink{0000-0002-3278-1433}  and Marko Poženel\,\orcidlink{0000-0002-1789-8668}}
  \IEEEauthorblockA{%
    Faculty of Computer and Information Science\\
    University of Ljubljana\\
    Ljubljana, Slovenia\\
    e-mail: {\tt$\lbrace$damjan.fujs\,|\,damjan.vavpotic|\,|\,tomaz.hovelja|\,marko.pozenel$\rbrace$@fri.uni-lj.si}
} }

\begin{document}
\maketitle

%#####################
% Abstract
%#####################

\begin{abstract}
This study investigates how access to Large Language Models (LLMs) and varying levels of professional software development experience affect the prioritization of cybersecurity requirements for web applications. Twenty-three postgraduate students participated in a research study to prioritize security requirements (SRs) using the MoSCoW method and subsequently rated their proposed solutions against multiple evaluation criteria. We divided participants into two groups (one with and the other without access to LLM support during the task). Results showed no significant differences related to LLM use, suggesting that access to LLMs did not noticeably influence how participants evaluated cybersecurity solutions. However, statistically significant differences emerged between experience groups for certain criteria, such as estimated cost to develop a feature, perceived impact on user experience, and risk assessment related to non-implementation of the proposed feature. Participants with more professional experience tended to provide higher ratings for user experience impact and lower risk estimates. 
\end{abstract}

%#####################
% Keywords
%#####################


%#####################
% Introduction
%#####################

\section{Introduction}

Software development is inherently dynamic, pushing organizations to adopt or tailor development methodologies to remain efficient and competitive \cite{mihelic2024}. Prioritization of cybersecurity requirements, especially when assisted by Large Language Models (LLMs) or shaped by prior experience, takes place within this evolving context, where structured yet adaptable decision-making is essential. Our study, therefore, addresses this crucial area. As systems grow increasingly complex and interconnected (as well as powered with Artificial Intelligence (AI) \cite{letier2025}), cybersecurity has become a critical concern that must be addressed early in the development lifecycle \cite{jungebloud2025}. Most decisions, we can argue, are probably still made by people. In practice, it is generally accepted that longer professional experience contributes to more effective decision-making. Also, in the literature, we can find some evidence to support such claims \cite{franke2015}. However, some emerging ideas \cite{kshetri2025} suggest that agentic AI systems could take on decision-making roles in specific areas of cybersecurity to address evolving cyber threats.

At this stage of the study, we focus on whether there are statistically significant differences in how selected Security Requirements (from here on referred to as SR/SRs as plural) are perceived by participants who used an LLM versus those who did not.  Specifically, we were interested in how participants estimated selected SRs across various evaluation criteria. This raises a broader and timely question: Can LLMs (or generative AI more broadly) begin to narrow or even erase the gap typically attributed to experience? The purpose of this paper is not to answer the question posed above, but to provide guidelines for further empirical research in the field of software engineering or software development. Additionally, we aimed to test the hypothesis on students, as they represent the next generation of software development professionals and are typically familiar with using LLMs.

Based on all the above, we hypothesize:
\begin{itemize}
    \item \textbf{H1}: Access to a LLM has a significant effect on how participants rate their proposed SRs across the given evaluation criteria.
    \item \textbf{H2}: Professional experience with software development has a significant effect on how participants rate their proposed SRs across the given evaluation criteria.
\end{itemize}

Based on the proposed hypotheses, our study offers two key contributions:

\begin{itemize}
    \item \textbf{C1}: Empirical insight into the limited impact of LLMs on cybersecurity decision-making among postgraduate students. The study provides evidence that LLM do not significantly influence how individuals prioritize or evaluate SRs.
    \item \textbf{C2}: Demonstration of the role of professional software development experience in prioritizing and evaluating SRs among postgraduate students. The study shows that professional software development experience significantly affects how students assess cost, user experience, and risk, highlighting the importance of practitioner expertise in shaping effective cybersecurity strategies.
\end{itemize}

The rest of the paper is organized as follows. Following this Introduction and Background section, Section \ref{relatedwork} briefly highlights existing related works. Section \ref{methodology} highlights the research methodology used. In Section \ref{results}, we present the results and discuss them briefly. In Section \ref{limitations} we point out the limitations of our study. Finally, the conclusion and future works are presented in Section \ref{conclusionFuture}.

%#####################
% Related Work
%#####################
\section{Related Work}
\label{relatedwork}

The creation of software requirements is a fundamental activity in any software project and is traditionally recognized as a labor-intensive, human-driven process \cite{Krishna2024}. Recent advances in AI, particularly the development of LLMs, have introduced new possibilities for supporting software engineering tasks such as SR engineering.

Prior research has explored various factors influencing the prioritization and evaluation of software and cybersecurity requirements, including tool support and individual expertise. Ronanki et al. \cite{ronanki2023} investigated the potential of ChatGPT to assist requirements elicitation. They found that requirements generated using ChatGPT were of higher quality than those generated by human requirements engineering experts. A similar observation was provided by Krishna et al. \cite{Krishna2024} where they found that LLMs can produce output comparable in quality to that of an entry-level software engineer when generating a software requirements specification. While general software requirements engineering has been extensively studied, particularly in terms of specification quality, tool support, and the role of human expertise, SR represents a specialized subset that introduces additional complexity.  For instance, in the study of Perry et al. \cite{Perry2023}, they found that participants who had access to an AI assistant wrote significantly less secure code than those without such support, raising concerns about overconfidence in automated tools in security-critical tasks. 

Moreover, previous study \cite{Jalali2019} did not find precise evidence that professional experience significantly shapes decision making in cybersecurity. In general, defining professional experience in software development is complex, as it encompasses diverse roles and learning paths, and it is similar in the field of cybersecurity. Baltes and Diehl \cite{Baltes2018} have shown that developers’ self-assessments of expertise are highly context-dependent. Vadlamani and Baysal \cite{Vadlamani2020} suggest, that while both knowledge and experience are necessary components of software development expertise, they are not sufficient on their own, as soft skills are also important.

The above mentioned studies highlight the role of AI tools and developer expertise in software engineering, yet little is known about how these factors influence the prioritization and evaluation of security requirements. This study addresses the gap by examining the combined effects of LLM access and professional experience on cybersecurity decision-making.

%#####################
% Research Methodology
%#####################

\section{Research Methodology}
\label{methodology}
We employed a controlled experiment  \cite{kampenes2009} in our research conducted in May 2025. The participants in the experiment were postgraduate students taking a course in Advanced software development methodologies, which is offered at the University of Ljubljana, Faculty of Computer and Information Science. The course is attended by students from technical disciplines who are enrolled in various master's programs, including Computer Science and Mathematics, Computer Science, and Multimedia. 

\begin{figure*}[h]
\includegraphics[width=1.0\linewidth]{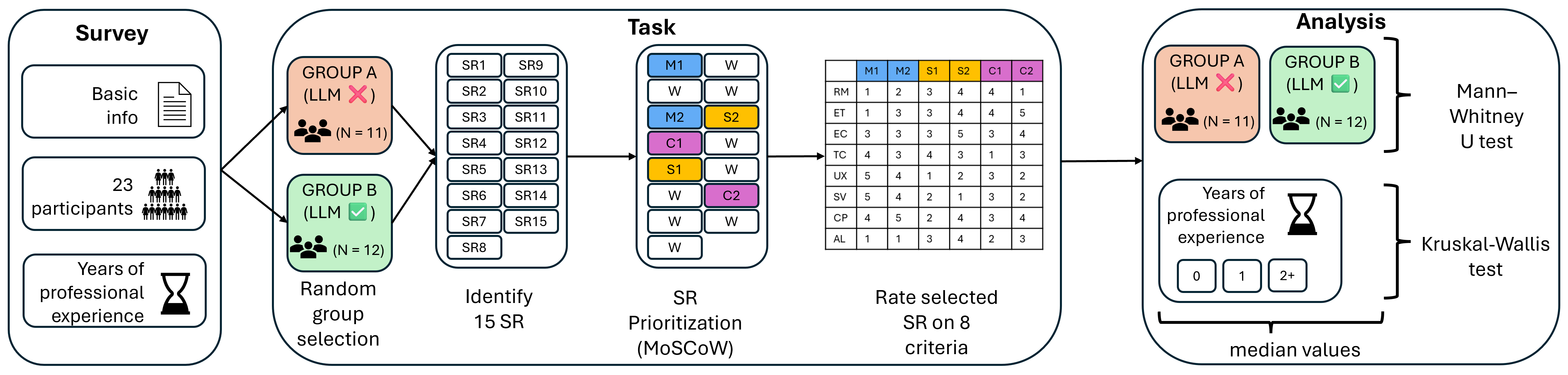}
\caption{The research framework consists of three main phases (e.g., Survey, Task and Analysis)}. Notes: LLM (Large Language Model), SR (Security Requirements).\label{approach}
\centering%
\end{figure*}

Figure \ref{approach} represents the entire research framework that consists of three main phases (e.g., Survey, Task and Analysis). The first phase involves conducting a survey. The second phase is an experiment in which participants complete a predefined task using a structured template. The final phase focuses on data analysis, including statistical testing to assess the significance between different groups and the reporting of median values.

In the first phase (\textit{Survey}), we received informed consent from the participants in the study, explained the course of the research to them, and gave them instructions. As part of the survey, in the first phase, we collected basic data about their studies and professional experience with software engineering. The exact question for years of professional experience with software engineering was: "Excluding education, how many years have you been 'professionally' involved in software development (e.g., student work, project work, etc.)?".

\captionsetup{font={footnotesize,sc},justification=centering,labelsep=period}%
\captionsetup{font={footnotesize,rm},justification=centering,labelsep=period}%

In the second phase (\textit{Task}), respondents were assigned to groups. Namely, $23$ research participants were divided into two groups; one group could use any LLM for the task (experimental group, N = 12), while the other could not (control group, N = 11). Both groups had the typical time available for practicals (i.e., 2 hours, including our instructions).

We prepared a scenario and a structured template for participants to enter their decisions into. The scenario was that, as part of their work on the project (as part of the course, they were developing software to support ScrumBan \cite{pozenel2023}), they were tasked with identifying $15$ SRs appropriate for enhancing the system's overall security posture. We limited participants to $15$ SRs in order to establish a unified framework while reflecting the resource constraints commonly encountered in industry settings. While developing the ScrumBan web application, students gained some experience with security aspects, particularly through implementing the login user story. The implementation of the login user story required them to handle authentication mechanisms, such as enforcing password policies (e.g., minimum length of 8 characters, inclusion of various character types and numbers). Additionally, they could improve the login procedure by implementing optional enhancements, such as a password strength meter or similar features, which further encourage consideration of usability and security.

The 15 SRs, initially identified by the participants, were subsequently prioritized using the MoSCoW method \cite{miranda2022}. The objective was to select $2X$ 'Must-have', $2X$ 'Should-have', and $2X$ 'Could-have' features from the set of 15. The remaining nine mechanisms were categorized as 'Won’t have this time'. A similar prioritization approach was used in Fujs et al. \cite{fujs2023}. The final step of the second task involved evaluating the six prioritized features using predefined criteria, as shown in Table \ref{criteria}.

\captionsetup{font={footnotesize,sc},justification=centering,labelsep=period}%
\begin{table*}[t]
\caption{The criterion with eight items by which respondents evaluated their selected priority security mechanisms.}\label{criteria}
\centering%
\footnotesize
\begin{tabular}{llp{5.0cm}p{6.9cm}}
\hline
\textit{ID} & \textit{Item} & \textit{1 - Lowest} & \textit{5 - Highest} \\
\hline
$1$ RM & Risk if not implemented & Minimal risk if not implemented & Critical security risk if not implemented\\
$2$ ET & Estimated time & Less than 1 hour & More than 10 hours\\
$3$ EC & Estimated cost & No cost, trivial to implement & High cost, external tools or experts needed\\
$4$ TC & Technical complexity & Very simple, can be done without research & Very complex, requires redesign or specialized knowledge\\
$5$ UX & UX impact & Almost no user impact on UX & High impact on user UX\\
$6$ SV & Security value & Adds minimal security benefit & Essential for application security\\
$7$ CP & Critical for production & Not needed for launch & Absolutely necessary before production release\\
$8$ AL & Abuse likelihood & Very unlikely to be abused & Very likely to be abused without this feature\\

\hline
\end{tabular}
\end{table*}
\captionsetup{font={footnotesize,rm},justification=centering,labelsep=period}%

As part of this step, we aimed at gathering additional quantitative data regarding the rationale behind the participants’ prioritization decisions. We used a $5$-point scale ranging from $1$ to $5$ for each evaluation criterion. For example, in the case of Estimated Time (ET), participants were asked to assess how long it would take to implement an overdue feature, where option/value $1$ corresponded to "less than $1$ hour" and option/value $5$ to "more than $10$ hours." Intermediate options (i.e., $2$, $3$, and $4$) were intentionally omitted to avoid over-constraining their responses and to encourage clearer distinctions in judgment. The study was conducted on-site at the university, allowing us to control whether participants were placed in a group with access to an LLM for the task or not. Additionally, we ensured that participants could ask questions if any part of the instructions was unclear.

In the last phase (\textit{Analysis}), we analyzed the collected data. We used appropriate non-parametric statistical tests \cite{okoye2024} given the sample size of $23$ respondents. Specifically, we employed the Mann-Whitney U test to assess whether there were statistically significant differences in prioritizations based on whether respondents did or did not use LLMs. Furthermore, respondents who were allowed to use LLMs had complete freedom to choose the LLM of their choice. Most chose the version of ChatGPT available at the time (N = 6), followed by DeepSeek (N = 2), Gemini (N = 2), Perplexity (N = 1), and Claude (N = 1). To examine differences across varying durations of professional experience, we used the Kruskal-Wallis test \cite{okoye2024}, suitable for comparing two or more groups. Based on these non-parametric tests, we then reported the Median. Based on their experience with professional software development, participants were divided into three groups: the first group included participants with zero years of experience (N = 10), the second group included participants with one year (N = 6), and the third group included participants with two or more years of experience (N = 7). This grouping was based on a qualitative judgment, as the participants were postgraduate students who were not yet formally employed. However, some had gained relevant professional software development experience through internships, freelance work, or other informal roles.

%#####################
% Results and Discussion
%#####################

\section{Results and Discussion}
\label{results}

Respondents selected up to six SRs using the MoSCoW prioritization method and subsequently rated each feature based on eight predefined criteria. This resulted in a total of 48 ratings per respondent (8 criteria × 6 prioritized SRs). An illustrative example of the rating form is shown in Figure \ref{approach} ("Rate selected SR on eight criteria"). 

The Mann-Whitney U test revealed no statistically significant differences across any of the evaluation criteria (column \textit{item} in Figure \ref{approach}). Based on these results, we conclude that access to an LLM did not significantly influence how respondents rated their proposed SRs. Therefore, Hypothesis H1 is not supported. Because we did not find significant differences, we do not report descriptive statistics (e.g., medians) for these comparisons. A possible explanation for the lack of statistically significant differences is that the LLM primarily served as a support tool for generating SRs, rather than influencing how participants evaluated their own solutions. Since the ratings were based on self-assessment, they were likely shaped more by the respondents’ individual understanding, confidence, or prior knowledge than by the presence or absence of the LLM. Furthermore, given that the participants were postgraduate students with limited formal industry experience, many may have lacked the expertise to critically evaluate the quality of their proposed SRs. As a result, their assessments may have been similar across groups, regardless of LLM access.

Pavlič et al. \cite{pavlic2024} studied user story effort estimation in agile environments, comparing development teams that had assistance in generative AI tools to control teams without such support (i.e., conventional effort estimation). Contrary to our findings, they found statistically significant differences between regular and AI-assisted teams. However, it is also worth noting that in our case, it is not the same problem domain, as our respondents evaluated their own SRs (based on eight criteria), while the study participants in Pavlič et al. \cite{pavlic2024} evaluated the effort in pre-prepared user stories. Moreover, it is important to take into account the fact that in our case, the use of LLM was an option for the experimental group (i.e., we did not force the experimental group to necessarily use LLM). We intended to create a setting that approximates real-world industry conditions, where access to a given technology, such as an LLM, is available. Still, its actual use remains at the discretion of the individual.

To test hypothesis H2, we conducted a Kruskal-Wallis test \cite{okoye2024}, which revealed statistically significant differences for specific evaluation items. Table \ref{rezultati} shows five items where statistically significant differences in scores occurred for certain prioritized SRs. Items for which no statistically significant differences have been found are not shown in Table \ref{rezultati} (there were $43$ such items). This result neither conclusively supports nor definitively refutes the hypothesis, as statistically significant differences were found for some items but not for most. However, it suggests that professional experience in software development may have an influence on certain evaluation criteria.

\captionsetup{font={footnotesize,sc},justification=centering,labelsep=period}
\begin{table}[htbp]
\caption{Median values for items by years of professional experience with software development. p-values indicate statistical significance for the item (ID).}\label{rezultati}
\centering
\footnotesize
\begin{tabular}{llll}
\hline
\textit{ID} & \textit{Years of professional experience} & \textit{Median} & \textit{p-value} \\
\hline
$S1EC$ & none ($0$)   & $2.00$ & $0.010$ \\
     & $1$ year       & $2.50$ &       \\
     & $2+$ years     & $3.00$ &       \\
$S1UX$ & none ($0$)   & $1.00$ & $0.036$ \\
     & $1$ year       & $1.00$ &       \\
     & $2+$ years     & $2.00$ &       \\
$S2RM$ & none ($0$)   & $4.00$ & $0.049$ \\
     & $1$ year       & $4.00$ &       \\
     & $2+$ years     & $3.00$ &       \\
$C1RM$ & none ($0$)   & $3.00$ & $0.003$ \\
     & $1$ year       & $2.00$ &       \\
     & $2+$ years     & $3.00$ &       \\
$C2EC$ & none ($0$)   & $1.50$ & $0.018$ \\
     & $1$ year       & $2.50$ &       \\
     & $2+$ years     & $2.00$ &       \\
\hline
\end{tabular}
\end{table}
\captionsetup{font={footnotesize,rm},justification=centering,labelsep=period}

The results indicate that statistically significant differences were found in the prioritization of should-have and could-have SRs, while no such differences were observed for must-have SRs. One possible explanation is that must-have SRs represent fundamental security mechanisms that are universally expected in any system (in addition, we also presented various cybersecurity mechanisms within the course, such as the OWASP (Open Worldwide Application Security Project) ASVS -  Application Security Verification Standard \cite{wen2023}). Additionally, participants may have based their decisions on the specific characteristics of the web application they developed, leading to more consistent prioritization in this category. 

C1RM achieved a p-value $< 0.01$, while S1EC, S1UX, S2RM and C2EC achieved a p-value $< 0.05$. In addition, it can also be observed that out of the eight criteria, statistically significant differences occur in three types, namely: Estimated Cost (EC), UX Impact (UX), and risk if not implemented (RM). Note that we were not interested in what actual SRs the respondents proposed, but rather in their values - that is, their assessments according to the criteria (see Table \ref{criteria}). Among these criteria, estimated cost (EC) stands out most prominently in both S1 and C2. The results show that participants without professional experience significantly underestimated the anticipated cost of developing a proposed feature. This could be due to limited exposure to real-world development constraints such as budgeting, resource allocation, or integration complexity. In contrast, more experienced participants likely drew from hands-on experience in estimating effort and understanding hidden development costs.  

In S1UX, participants with two or more years of professional experience stand out by assigning a higher median rating to the impact of the proposed feature on user experience. Similarly, in S2RM, participants with two or more years of professional experience provided slightly lower median estimates of the risk associated with not implementing the proposed feature, compared to those with no experience or only one year of experience. 

%#####################
% Limitations
%#####################

\section{Limitations}
\label{limitations}
While we can see some differences, it is difficult to argue about the influence of professional software development experience and the use of LLM based on these results alone. Thus, some limitations should be considered in the interpretation of these findings. First, the number of respondents is relatively small, limiting the findings' statistical power and generalizability. Second, although certain trends emerge, for instance, more experienced participants assigning higher user experience impact or lower risk estimates, these differences may also reflect individual interpretation or subjective biases rather than consistent effects of professional experience. Third, the ratings are self-reported, and participants may have relied on intuition or heuristics rather than systematic analysis, further complicating the interpretation. Therefore, while the data suggest a potential link between experience and how participants assess different aspects of cybersecurity features, these observations should be interpreted with caution.

Fourth, a potential selection bias may have occurred during group selection, as participants were assigned based on their position within the computer classroom (we counted and placed the first $11$ individuals present in one group and the remaining $12$ in another). This method may have unintentionally clustered individuals with similar characteristics, such as higher academic achievement, thereby affecting group comparability. 

Fifth, another limitation concerns the nature and depth of LLM integration. Participants may not have fully utilized the LLM's capabilities due to time constraints, unfamiliarity with prompting, or skepticism about the tool's relevance, etc.

%#####################
% Conclusion and Future Works
%#####################

\section{Conclusion and Future Works}
\label{conclusionFuture}

In our research, $23$ postgraduate students took part in a study aimed at prioritizing SRs using the MoSCoW method. Afterward, they evaluated their proposed solutions against several criteria. The participants were split into two groups: one had access to LLM support during the task, while the other did not.

The study found that access to LLM did not significantly influence how participants prioritized SRs. However, professional software development experience played a notable role in shaping evaluations. Participants with more experience rated the impact on user experience higher and perceived lower risks associated with not implementing certain features. Significant differences were also observed with estimated cost, user experience, and risk assessment, highlighting the importance of domain expertise in cybersecurity decision-making.

While the current study provides valuable insights into the use of LLMs for evaluation tasks, several opportunities remain for further exploration. Future research should consider designing tasks that require deeper interaction with the model to better evaluate its potential impact for "evaluation tasks". 

Future studies could incorporate external expert evaluations or peer reviews to obtain more objective assessments of solution quality. For example, it would also make sense to look at the quality - what SRs they have identified and how they have prioritized them (what mechanisms are there, which vulnerabilities do they cover, etc.). Moreover, future research could also explore how different professional roles interact with and evaluate model outputs. For instance, developers may focus on technical accuracy and implementation feasibility, project managers on delivery timelines and resource constraints, and stakeholders on strategic value and return on investment.

%#####################
% Acknowledgments
%#####################

\section*{Acknowledgments}
This work was funded by the Slovenian Research and Innovation Agency [Grant Number P2–0426]. The authors thank the participants in the research for their eager collaboration.

%#####################
% References
%#####################

\printbibliography

@article{fujs2023,
  title={Balancing software and training requirements for information security},
  author={Fujs, Damjan and Vrhovec, Simon and Vavpoti{\v{c}}, Damjan},
  journal={Computers \& security},
  volume={134},
  pages={103467},
  year={2023},
  publisher={Elsevier}
}

@article{kampenes2009,
  title={A systematic review of quasi-experiments in software engineering},
  author={Kampenes, Vigdis By and Dyb{\aa}, Tore and Hannay, Jo E and Sj{\o}berg, Dag IK},
  journal={Information and Software Technology},
  volume={51},
  number={1},
  pages={71--82},
  year={2009},
  publisher={Elsevier}
}

@inproceedings{miranda2022,
  title={Moscow rules: A quantitative expos{\'e}},
  author={Miranda, Eduardo},
  booktitle={International Conference on Agile Software Development},
  pages={19--34},
  year={2022},
  organization={Springer}
}

@incollection{okoye2024,
  title={Mann--Whitney U Test and Kruskal--Wallis H Test Statistics in R},
  author={Okoye, Kingsley and Hosseini, Samira},
  booktitle={R programming: Statistical data analysis in research},
  pages={225--246},
  year={2024},
  publisher={Springer}
}

@article{pavlic2024,
  title={Can Large-Language Models Replace Humans in Agile Effort Estimation? Lessons from a Controlled Experiment},
  author={Pavli{\v{c}}, Luka and Saklamaeva, Vasilka and Berani{\v{c}}, Tina},
  journal={Applied Sciences},
  volume={14},
  number={24},
  pages={12006},
  year={2024},
  publisher={MDPI}
}

@article{wen2023,
  title={A quantitative security evaluation and analysis model for web applications based on OWASP application security verification standard},
  author={Wen, Shao-Fang and Katt, Basel},
  journal={Computers \& Security},
  volume={135},
  pages={103532},
  year={2023},
  publisher={Elsevier}
}

@article{mihelic2024,
  title={Delegation-based agile secure software development approach for small and medium-sized businesses},
  author={Miheli{\v{c}}, An{\v{z}}e and Vrhovec, Simon and Markelj, Bla{\v{z}} and Hovelja, Toma{\v{z}}},
  journal={IEEE Access},
  year={2024},
  publisher={IEEE},
  pages={189611-189635}
}

@article{kshetri2025,
  title={Transforming cybersecurity with agentic AI to combat emerging cyber threats},
  author={Kshetri, Nir},
  journal={Telecommunications Policy},
  pages={102976},
  year={2025},
  publisher={Elsevier}
}

@article{letier2025,
  title={Obstacle Analysis in Requirements Engineering: Retrospective and Emerging Challenges},
  author={Letier, Emmanuel and Van Lamsweerde, Axel},
  journal={IEEE Transactions on Software Engineering},
  year={2025},
  publisher={IEEE},
  pages={795-801}
}

@article{jungebloud2025,
  title={Model-based structural and behavioural cybersecurity risk assessment in system designs},
  author={Jungebloud, Tino and Nguyen, Nhung H and Kim, Dan Dongseong and Zimmermann, Armin},
  journal={Computers \& Security},
  pages={104543},
  year={2025},
  publisher={Elsevier}
}

@article{franke2015,
  title={Experimental evidence on decision-making in availability service level agreements},
  author={Franke, Ulrik and Buschle, Markus},
  journal={IEEE Transactions on Network and Service Management},
  volume={13},
  number={1},
  pages={58--70},
  year={2015},
  publisher={IEEE}
}

@inproceedings{ronanki2023,
   author={Ronanki, Krishna and Berger, Christian and Horkoff, Jennifer},
   title={Investigating ChatGPT’s Potential to Assist in Requirements Elicitation Processes},
   booktitle={2023 49th Euromicro Conference on Software Engineering and Advanced Applications (SEAA)},
   month={9},
   pages={354-361},
   publisher={IEEE},
   year={2023}
}

@inproceedings{Krishna2024,
   author={Madhava Krishna and Bhagesh Gaur and Arsh Verma and Pankaj Jalote},
   booktitle={2024 IEEE 32nd International Requirements Engineering Conference (RE)},
   month={6},
   pages={475-483},
   publisher={IEEE},
   title={Using LLMs in Software Requirements Specifications: An Empirical Evaluation},
   year={2024}
}

@article{Perry2023,
   author = {Neil Perry and Megha Srivastava and Deepak Kumar and Dan Boneh},
   journal = {CCS 2023 - Proceedings of the 2023 ACM SIGSAC Conference on Computer and Communications Security},
   month = {12},
   pages = {2785-2799},
   publisher = {Association for Computing Machinery, Inc},
   title = {Do Users Write More Insecure Code with AI Assistants?},
   year = {2023}
}

@article{Jalali2019,
   author = {Mohammad S. Jalali and Michael Siegel and Stuart Madnick},
   issue = {1},
   journal = {The Journal of Strategic Information Systems},
   month = {3},
   pages = {66-82},
   publisher = {Elsevier B.V.},
   title = {Decision-making and biases in cybersecurity capability development: Evidence from a simulation game experiment},
   volume = {28},
   year = {2019}
}

@inproceedings{Baltes2018,
   author = {Sebastian Baltes and Stephan Diehl},
   city = {New York, NY, USA},
   isbn = {9781450355735},
   booktitle = {Proceedings of the 2018 26th ACM Joint Meeting on European Software Engineering Conference and Symposium on the Foundations of Software Engineering},
   month = {10},
   pages = {187-200},
   publisher = {ACM},
   title = {Towards a theory of software development expertise},
   year = {2018}
}

@inproceedings{Vadlamani2020,
   author = {Sri Lakshmi Vadlamani and Olga Baysal},
   booktitle = {2020 IEEE International Conference on Software Maintenance and Evolution (ICSME)},
   month = {9},
   pages = {312-323},
   publisher = {IEEE},
   title = {Studying Software Developer Expertise and Contributions in Stack Overflow and GitHub},
   year = {2020}
}

@article{pozenel2023,
  title={Agile effort estimation: Comparing the accuracy and efficiency of planning poker, bucket system, and affinity estimation methods},
  author={Po{\v{z}}enel, Marko and F{\"u}rst, Luka and Vavpoti{\v{c}}, Damjan and Hovelja, Toma{\v{z}}},
  journal={International Journal of Software Engineering and Knowledge Engineering},
  volume={33},
  number={11n12},
  pages={1923--1950},
  year={2023},
  publisher={World Scientific}
}
\end{document}